# Integrated Electro-Optic Isolator on Thin Film Lithium Niobate

Mengjie Yu[1,2,*], Rebecca Cheng[1,*], Christian Reimer[3], Lingyan He[3], Kevin Luke[3], Eric Puma[1,4], Linbo Shao[1], Amirhassan Shams-Ansari[1], Hannah R. Grant[5], Leif Johansson[5], Mian Zhang[3], and Marko Lončar[1]

[1]John A. Paulson School of Engineering and Applied Sciences, Harvard University, Cambridge, MA 02138
[2]Ming Hsieh Department of Electrical and Computer Engineering, University of Southern California, Los Angeles, CA 90089, USA
[3]HyperLight, 501 Massachusetts Avenue, Cambridge, MA 02139
[4]Draper Scholar, The Charles Stark Draper Laboratory, Inc., 555 Technology Square, Cambridge, Massachusetts 02139, USA
[5]Freedom Photonics, 41 Aero Camino, Goleta CA, USA
*These authors contributed equally; * Corresponding author: loncar@seas.harvard.edu

**Optical isolator is an indispensable component of almost any optical system and is used to protect a laser from unwanted reflections for phase-stable coherent operation. The development of chip-scale optical systems, powered by semiconductor lasers integrated on the same chip, has resulted in a need for a fully integrated optical isolator. However, conventional approaches based on application of magneto-optic materials to break the reciprocity and provide required isolation have significant challenges in terms of material processing and insertion loss. As a result, many magnetic-free approaches have been explored, including acousto-optics, optical nonlinearity, and electro-optics. However, to date, the realization of an integrated isolator with low insertion loss, high isolation ratio, broad bandwidth, and low power consumption on a monolithic material platform is still absent. Here we realize non-reciprocal traveling-wave EO-based isolator on thin-film LN, enabling maximum optical isolation of 48 dB and an on-chip insertion loss of 0.5 dB using a single-frequency microwave drive at 21-dBm RF power. The isolation ratio is verified to be larger than 37 dB across a tunable optical wavelength range from 1510 to 1630 nm. We verify that our hybrid DFB laser - LN isolator module successfully protects the single-mode operation and the linewidth of the DFB laser from reflection. Our result is a significant step towards a practical high-performance optical isolator on chip.**

Recent decades have witnessed the rapid advancement of integrated photonic technology for applications including optical communications[1,2], microwave photonics[3], computing[4,5], optical atomic clock[6], quantum information processing[7], light detection and ranging[8] and sensing[9,10]. In parallel, Hertz-linewidth semiconductor lasers have been recently demonstrated on chip[11], showing promise as photonic engines for future large-scale photonic processors. Therefore, the urgent need to combine the narrow-linewidth semiconductor laser with the functional photonic modules demands a high-performance reliable integrated optical isolator.

Various approaches for realizing integrated optical isolators have been demonstrated, including magneto-optics[12–16], electro-optics[17–23], acousto-optics[24–26] as well as optical nonlinearities[27–30]. Magneto-optical isolator features robust operation and broad bandwidth but faces significant challenges of unconventional material integration, external magnetic fields and high insertion loss. Nonlinear optical approaches either require additional optical pump lasers and filtering[29,30] to remove the pump or show power-dependent isolation[27,28]. Recent demonstrations based on electro-optics and acousto-optics require narrowband optical cavities[23–25] to enhance non-reciprocity. engineered phase-matching of two optical modes[25,26] or hybrid integration of piezoelectric material and silicon photonics[24,26]. These limit the operational bandwidth and system scalability. Electro-optic (EO) waveguide-based device has immense potential to overcome these limitation. However, previous demonstrations based on silicon or III-V EO modulators[20–22] have suffered from high insertion loss and low isolation ratio due to the tradeoff between phase modulation and absorption. Therefore, realization of a high performance integrated isolator with low insertion loss, high isolation ratio, low power consumption, broad bandwidth and high fabrication tolerance remains challenging.

Thin film (TF) lithium niobate (LN) is well positioned to address this challenge, as it simultaneously supports phase-only modulation and low propagation loss[31]. The recent advancement of integrated LN platform has enabled ultra low-loss travelling-wave EO modulators operating at CMOS-compatible voltages and high EO bandwidth[31]. Other excellent photonic properties[32], including wide transparency window, large second order and Kerr nonlinearity and piezoelectric coefficient, makes LN a highly compelling choice for realizing future coherent optical systems.

In this paper, we demonstrate an integrated isolator based on a travelling-wave-based phase modulator realized in TFLN, and use it to protect distributed feedback (DFB) laser edge-coupled to the TFLN chip (Figure 1). We use single-tone sinusoidal microwave field, at the frequency of $f_{RF}$, propagating in the direction opposite from the optical signal. The total phase accumulation $\phi(f_{RF}, t)$ is equal to $\frac{V_{pk}}{V_{\pi}L} \pi \int_0^L e^{-\alpha z/2} \cos\left[2\pi f_{RF}\left(\frac{z}{c} n_{opt} \pm \frac{z}{c} n_{RF} - t\right)\right] dz$, where $V_{pk}$, $V_{\pi}$, $\alpha$, $n_{opt}$, $n_{RF}$, and c is peak voltage, half-wave voltage, microwave power attenuation constant, optical group index, microwave phase index and light velocity in vacuum, respectively. + and – are chosen to indicate situations when the optical and microwave fields are counter-propagating and co-propagating, respectively. Since the laser output is counter-propagating with the microwave signal, the phase accumulation is a periodic function of the propagating distance in the modulator and equals zero when the RF driving frequency $f_{RF} = mc/(2Ln_{opt})$ where m is an integer number, assuming $\alpha$ = 0. In this case, the laser light is not modulated and is fully transmitted in the forward direction. On the other hand, reflected light travelling in the backward direction co-propagates with the microwave field, resulting in a linearly increasing phase accumulation for any given $t$. In this case, the reflected light at the input frequency $f_0$ can be fully depleted when the modulation depth $\beta \approx 0.765\pi$ satisfying $J_0(\beta) = 0$ where $J_0$ is the 0th order Bessel function. All the optical power is distributed to frequencies equal to $f_0 + nf_{RF}$, where n is an integer (Fig.1b).

Figure 1c shows the optical micrograph of the integrated LN chip with modulator lengths of 1.25 cm to 2 cm at a step of 0.25 cm. The chip is fabricated on a 600-nm X-cut LN wafer with an etch depth of 320 nm (see Methods). The group velocity of the light matches the phase velocity of the microwave ($n_{opt} = n_{RF} = 2.26$) to ensure the efficient phase modulation at microwave driving frequencies at tens of GHz. The chip has both input and output mode converters[33] for a low-loss interface with lensed fiber (3-dB/facet IL) and DFB laser (2-dB/facet IL). Our device utilizes a simple optical structure of a single modulated waveguide and operates with a single-frequency microwave drive given that the microwave power is tuned to the pump depletion point.

Next, we show that the integrated phase modulator allows for high performance optical isolation at the laser frequency. First, we characterize the isolation ratio and optical bandwidth using a tunable continuous-wave laser (see Methods) and a phase modulator of 1.75 cm. The isolation ratio is defined as the ratio of optical power propagating in the forward and backward direction at the laser frequency $f_0$. Figure 2a plots the optical spectrum in both forward and backward directions using $f_{RF} = 26.5$ GHz (m = 7). In the forward direction, the laser light (wavelength of 1553.2 nm) is nearly completely transmitted, with an on-chip insertion loss of 0.5 dB. In the backward direction, however, most of the light is coupled to sidebands, at a frequency spacing equal to harmonics of the $f_{RF}$ (= 26.5 GHz). Next, we demonstrate the ability of our isolator to operate at different microwave frequencies, $f_{RF}$ of 26.5 and 10.4 GHz (Figure 2b). At 26.5 GHz, an isolation ratio of 48 dB is achieved at the pump wavelength, using 21 dBm of microwave power, which corresponds to $\beta \approx 0.765\pi$ with a $V_\pi = 4.7$ V. Since the microwave power scales as $P \sim (0.0765 \times V_\pi)^2$, power reductions are possible by reducing $V_\pi$. At 10.4 GHz, an isolation ratio is estimated to be > 44 dB (measurement limited by the instrumental resolution). At both frequencies, the isolation is likely limited by the intensity noise of the RF signal and the purity of the light polarization. The ability to achieve large isolation using different RF frequencies offers flexibility in the system design: for example it allows sidebands to be placed off-resonance with the laser cavity, or outside the laser instability range, thus ensuring stable laser operation. We note that the forward light exhibits the first sideband that are 37 dB below the pump, which is attributed to the microwave power attenuation along the electrode ($\alpha = 0.7$ dB/cm/GHz$^{0.5}$). These sidebands cause minimal loss to the laser signal and could be filtered out . We further tested the optical bandwidth of our device by tuning the laser wavelength from 1510 to 1630 nm. As shown in Fig. 2c, the optical isolation remains >37 dB across the entire tuning range. We note that this bandwidth is tested using a single-frequency laser and should not apply for broadband light source input.

In cases where the total power reflected to the laser needs to be minimized, add-drop ring-resonator filter, transmitting at laser wavelength and featuring free spectral range (FSR) larger than $f_{RF}$, can be combined with the phase modulator, as shown in Fig. 3a. Here, the laser light first enters the ring at Port 1 and exists through Port 2 and then goes through phase modulator without experiencing modulation. The reflected light goes through phase modulator where sidebands are generated. The sidebands are off-resonance with the filter, do not couple to it, and are directed towards the Port 3. The reflected power can be easily suppressed since the sidebands are multiple $f_{RF}$ away from the laser frequency. The free spectral range (FSR) and the intrinsic $Q$ factor of the microring resonator is 120 GHz and $1 \times 10^6$, respectively. Figure 3b shows the measured optical spectra, where the highest sideband power is suppressed and is 34 dB below the laser power and the 48-dB isolation ratio remains unchanged. The suppression can be further increased using a higher $Q$ factor microresonator and a higher $f_{RF}$. The total backward-propagating power is 30dB lower than the forward-propagating power . In order to have wide tuning range of optical wavelengths, an integrated NiCr heater is fabricated (Figure 3c). The tuning efficiency is measured to be 8nm/W. Thus, < 125 mW of heater power is required to tune the filter across one FSR of 120 GHz, therefore covering the entire wavelength range. Figure 3d shows the relationship between the insertion loss induced by the filter and the sideband suppression ratio based on the intrinsic $Q$ factor. We measured an insertion loss of 0.9 dB and suppression ratio of 34 dB, which is in good agreement with the simulation.

Finally, we verify the performance of our optical isolator using a complete III-V-LN laser module, as shown in Fig. 1a. We construct the system by edge-coupling a prefabricated InP distributed feedback (DFB) laser with the LN isolator chip (Fig. 4a). The coupling between the III-V and LN components is carefully engineered via mode-matching, which leads to a measured 2-dB loss at the interface (see Methods). To assess the effectiveness of the isolator in protecting the DFB laser, we send the output of the circuit through a 90:10 beamsplitter. The 90% port gets sent to a retroreflector which can be turned on/off to act as unwanted facet reflections at high laser power operation. The 10% port is sent to a self-heterodyne measurement system to monitor the DFB linewidth at different operating conditions (Fig. 4b). Fig. 4c shows the linewidth measurement at different laser operating conditions. We see that when high-power reflection is introduced with the isolator turned off, multimode oscilation is observed with additional peaks in the spectrum spaced ~6.5 MHz from the center peak. This frequency spacing corresponds to the FSR of the cavity formed by the retroreflector and DFB laser. We also verified that the relative intensity noise out of the entire system is similar between when the isolator is turned on and off (see Methods). We posit that there is additional noise from facet reflections on the LN chip, adding further noise to the measurement, which we believe explains the asymmetric spacing of the linewidth peaks. Additional traces of the intensity measurement illustrating this noise are given in the Methods. When the LN isolator is turned on, the linewidth of the laser is restored, showing that our optical isolator successfully protects the single mode operation of laser from reflection. To our knowledge, this is the first demonstration of an on-chip laser and isolator module for phase stablity protection, which clearly indicates the practicality of our isolator approach.

In conclusion, we demonstrate an integrated electro-optic modulator-based optical isolator on thin film lithium niobate, and verify its ability to protect the phase stability of an edge-coupled on-chip laser. Our device features 48 dB isolation ratio at the laser frequency, 30-dB power isolation, 0.5 dB on-chip insertion loss (1.5 dB with the ring filter) and a tunable optical operating wavelength across C and L bands, using 21-dBm microwave power. To the best of our knowledge, this is the largest optical isolation and the lowest on-chip insertion loss ever demonstrated for an integrated optical isolators. The non-resonant EO device is compatible with a wide range of laser wavelengths. The isolator only requires a counter-propagating single-tone microwave signal and is independent of input laser power or phase. We overcome the limitation of high insertion loss and low isolation contrast which typically exisits in a non-resonant system due to lack of Purcell enhancement factor in the light-matter interaction[25,26]. Our approach signifcantly benefits from the combination of low insertion loss and high EO efficiency of the integrated LN platform and offers the highly competitive performance across all metrics compared to other on-chip approaches (see Methods). We envision further reduction of both the on-chip modulator and filter insertion loss to less than 0.1 dB based on the state-of-art linear loss of 3dB/m on thin film LN[34] and of laser-LN chip coupling loss to <2 dB through implementation of an anti-reflection coating on the LN side and direct heterogeneous or homogeous integration of the laser and LN device to reduce the coupling gap. Further improvement of power isolation to 60 dB can be achieved by cascading another tunabel ring-based filter. Broadband optical isolation can be achieved based on interference effect by splitting into multiple modulation paths[22]. Microwave power consumption can be reduced to 30 mW by reducing the half-wave voltage by half via implementing a double-pass phase modulator design[35] and further down to 13 mW by ultizing high performance electrode design[36] with 0.25 dB/cm/GHz$^{0.5}$. The microwave power reduction to tens of mW could avoid the usage of a microwave amplifier and make the isolator directly driven by a low-noise microwave synthesizer, which will continue to improve the isolation ratio closer to the theroretical limit of near infinity. At last, the entire laser module can be put on the same substrate via hybrid integration between a semiconductor laser and the LN isolator[37,38]. We envision a fully functional coherent on-chip laser system in the future for applications from optical communication to time-frequency transfer.

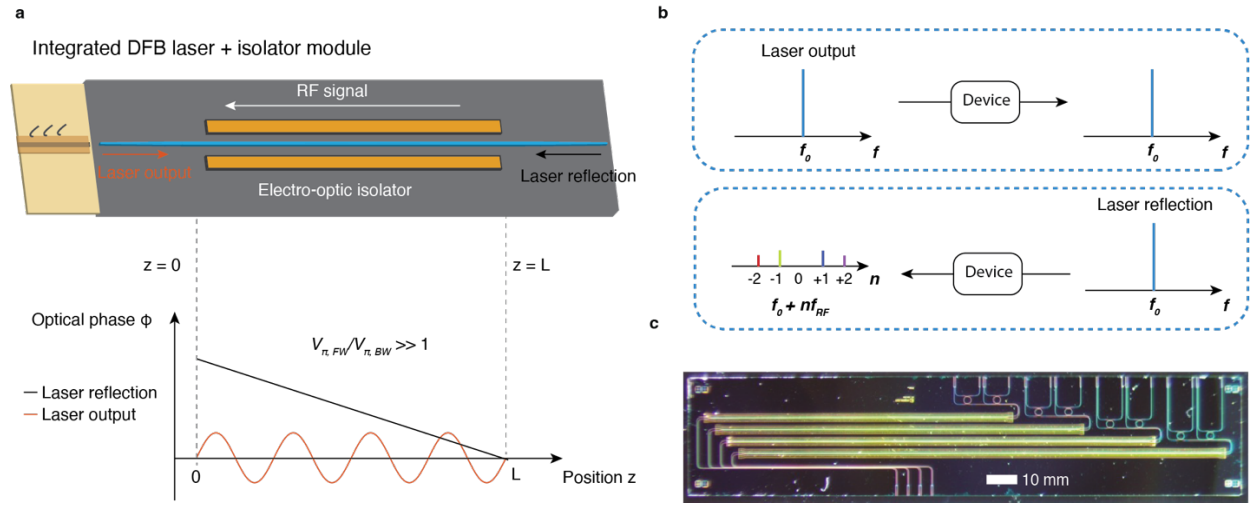

**Fig. 1 | Concept of an integrated electro-optic isolator. a,** The schematic of an integrated laser system consisting of a III-V-based distributed feedback laser (DFB) followed by an electro-optic isolator on thin film lithium niobate (LN). The microwave signal is counter-propagating with respect to the DFB laser output. Under single-tone travelling-wave-based microwave drive at $f_{RF}$, the accumulated optical phase of the laser output after passing a phase modulator of length $L$ is equal to zero when $f_{RF} = mc/(2Ln_{opt})$ where $m$ is an integer number and $c$ and $n_{opt}$ are light velocity and optical group index, respectively. Any reflected laser light which propagates in the backward direction will undergo efficient phase modulation, which results in $\frac{V_{\pi,FW}}{V_{\pi,BW}} \gg 1$. FW: forward; BW: backward. **b,** The illustration of the travelling-wave EO isolator in frequency domain. The laser output at frequency $f_0$ is fully transmitted since there is no added phase modulation. All the energy of the reflection signal at frequency $f_0$ is distributed to the optical sidebands at frequency equal to $f_0 + nf_{RF}$, where n is an integer number. The energy at frequency $f_0$ can be fully depleted when the modulation depth $\beta \approx 0.765\pi$ satisfying $J_0(\beta) = 0$. **c,** The optical micrograph of the EO isolator chip on thin film LN, which consists of four individual devices with varying modulation length $L$ of 1.25, 1.5, 1.75 and 2cm.

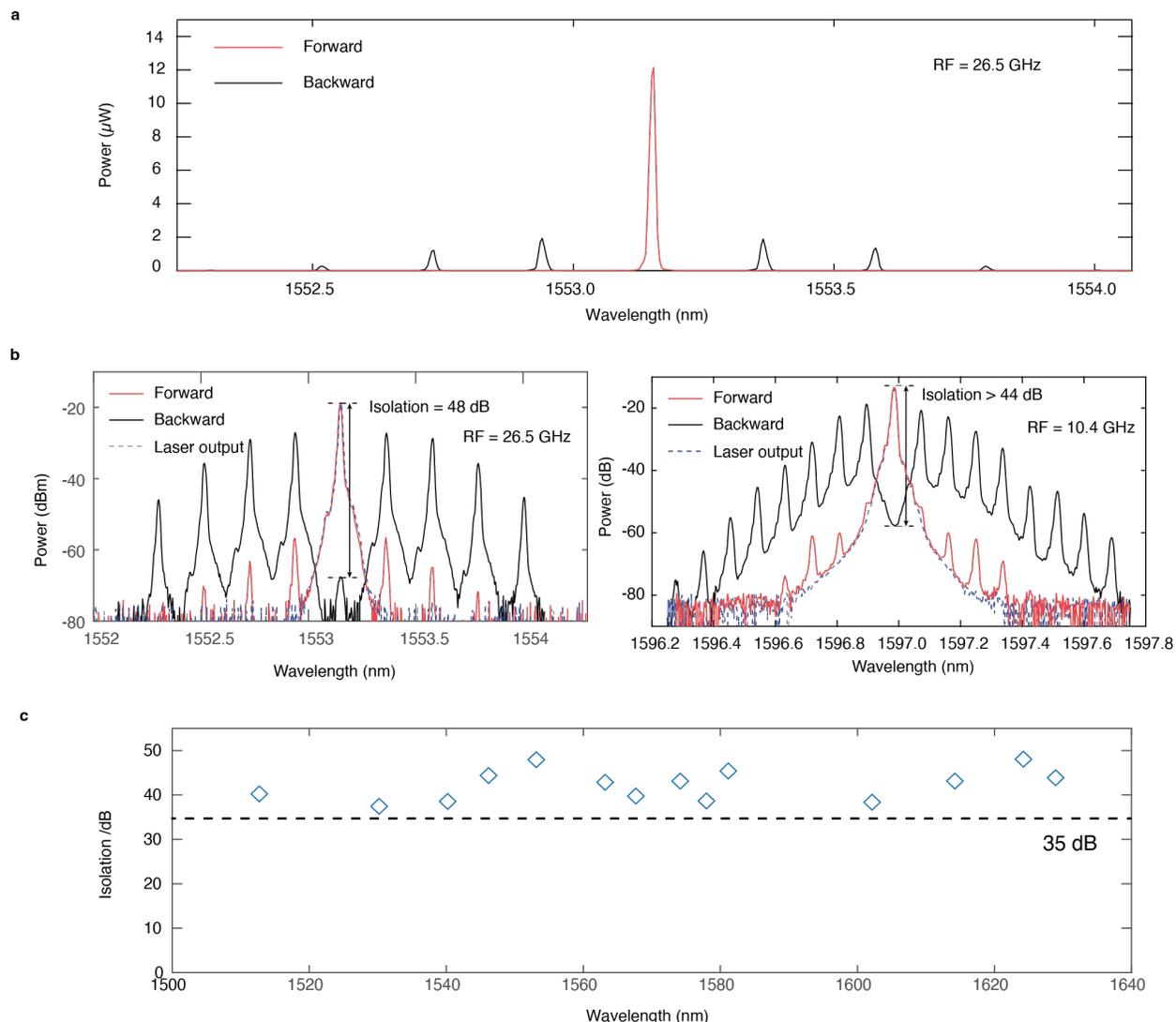

**Fig. 2 | Characterization of isolation and optical bandwidth of the electro-optic isolator.**
**a,** The optical spectrum of the forward and backward optical signal when the microwave signal is propagating in the backward direction at a frequency of 26.5 GHz and a power of 21 dBm (125 mW). **b,** The optical spectra (power in dB scale) at two different microwave driving frequencies of 26.5 GHz and 10.4 GHz, respectively. The achieved isolation at the input laser frequency is 48 dB at 26.5 GHz driving, and > 44 dB at 10.4 GHz driving (instrumental resolution limited), respectively. We have observed sidebands with > 37 dB lower power than the pump line in the forward signal in both cases. On-chip insertion loss is 0.5 dB. **c,** The isolation ratio as a function of optical wavelength. The isolation ratio remains larger than 37 dB across a tuning range from 1510 – 1630 nm.

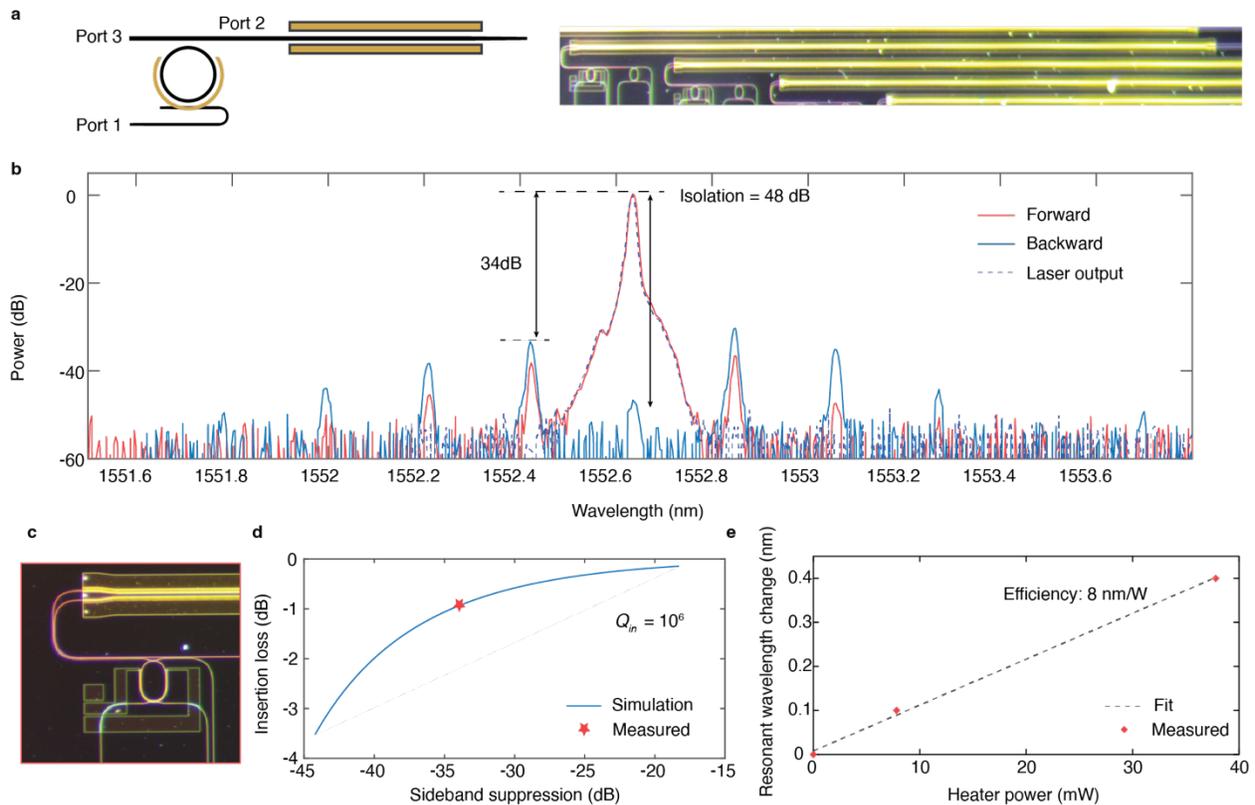

**Fig. 3 | Integration of a tunable add-drop ring filter on one thin film lithium niobate chip. a,** The schematic of an add-drop-ring filter with an integrated heater, followed by a travelling-wave EO isolator**.** The optical micrograph of the device is shown (right). **b,** The optical spectra of the forward and backward signals. Laser output spectrum is taken when the microwave drive is turned off. With the ring filter, the sideband power is 34 dB lower (0.04%) as compared to the pump power while the isolation ratio remains 48 dB. **c,** The optical micrograph of the ring filter with an integrated NiCr heater**. d,** The insertion loss as a function of sideband suppression. The sideband suppression is defined as the power ratio between the first sideband and the pump. We measured an insertion loss of 0.9 dB along with the 34-dB sideband suppression at 26.5 GHz. **e,** Characterization of the integrated heater. The resonance wavelength is recorded as a function of heater power. We can tune the cavity resonance over one free spectral range of 1.0 nm with 125mW heater power, based on a fitted tuning efficiency of 8nm/W.

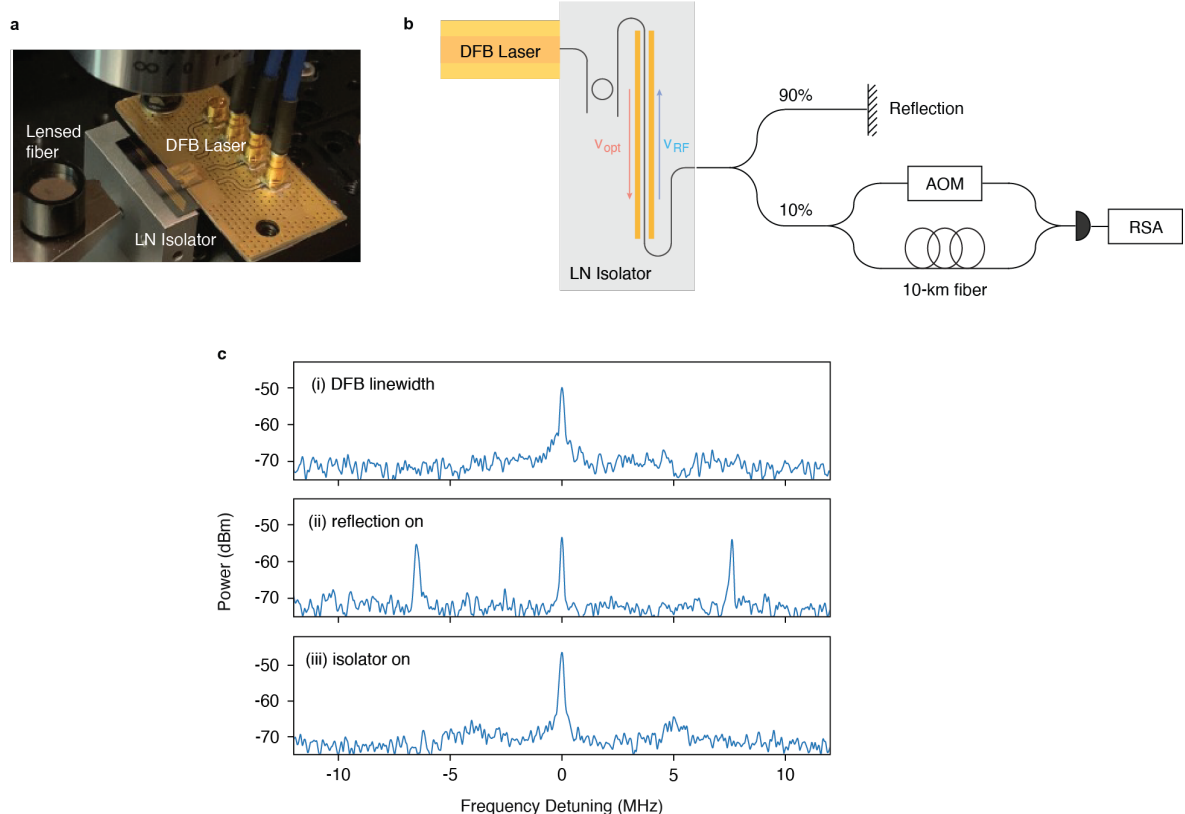

**Fig. 4 | Characterization of the DFB laser and isolator module. a,** Photograph of measurement system. An InP DFB laser is butt-coupled to the LN isolator chip. The laser system output is collected with a lensed fiber; **b)** Measurement setup. The output of the laser system is split between a reflection port and a self-heterodyne linewidth measurement system to monitor the DFB linewidth. AOM: acousto-optic modulator, RSA: RF spectrum analyzer; **c)** Linewidth measurements are given for (i) the laser module alone, (ii) with strong reflection and isolator turned off, (iii) with strong reflection and isolator turned on. RSA resolution bandwidth: 20 kHz.

## Methods
### Device fabrication.

Devices are fabricated on thin-film X-cut lithium niobate on insulator wafer (NanoLN) with 600 nm film thickness and an buried oxide thickness of 2 μm. Electron-beam lithography (EBL) is used to define the optical waveguide with hydrogen silsesquioxane resist, which is then partially etched by Ar+ -based reactive ion etching with an etch depth of 320 nm. The entire device is cladded with silicon dioxide via plasma-enhanced chemical vapor deposition (PECVD). Windows for the mode size converter are defined using photolithography and hydrofluoric acid wet etching of the $SiO_2$ cladding. To define the tip of the mode converter (Extended Data Fig. 2), an additional EBL and Ar etching step is done to etch through the remaining 280 nm of slab. The entire device is then cladded a second time with PECVD oxide. The microwave electrodes (1.6 μm Au) are defined using electron-beam lithography and photolithography and metallized using electron beam evaporation. The NiCr heaters are defined using photolithography and also metallized via electron-beam evaporation. The facets of the LN chip are etched with deep reactive ion etching.

### Experimental Setup.

The isolation and sideband suppression of our device are evaluated using a tunable laser (Santec-710) and a 2x2 switch to switch the direction of the light coupled onto the LN chip. Polarization controllers are used in both the forward and backward direction to ensure optimal operation. An optical spectrum analyzer (OSA) is used to collect the spectra first without isolation, and in the forward and backward direction with the isolator turned on. The phase modulator electrodes are driven by GS probes with signal generator. If using the add-drop filter, the resonance of the filter ring is thermally tuned to the wavelength of the laser.

The effectiveness of the isolator in protecting a DFB laser under strong reflection is measured by replacing the lensed fiber input of the isolator with a III-V DFB laser. The lasing wavelength of the laser is tuned via operating current to match the add-drop filter. The output of the laser module is split between a retroreflector and linewidth and relative intensity noise measurement. We estimate the total fiber length between the LN chip and the retroreflector to be ~16 m, corresponding to a cavity FSR of ~6.5 MHz.

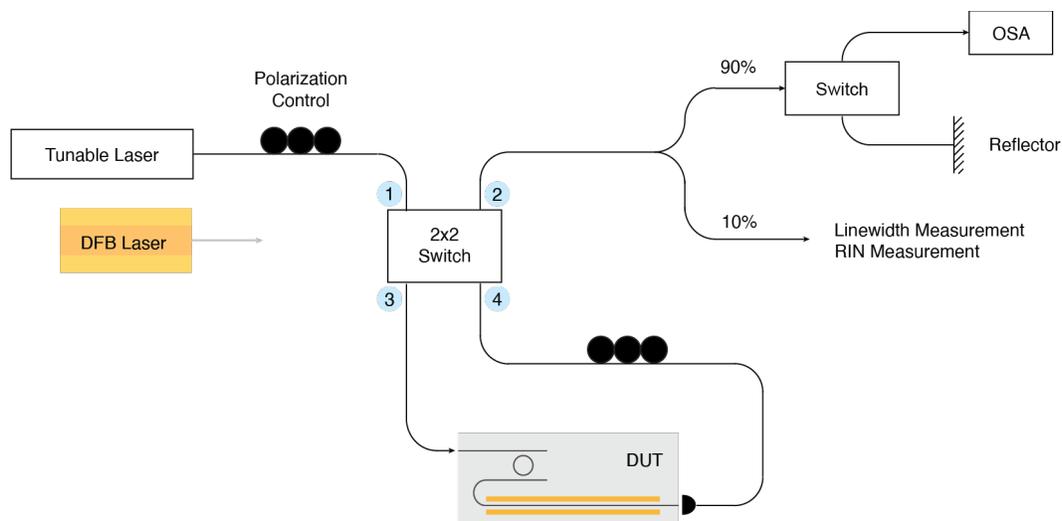

**Extended Data Fig. 1 | Full measurement setup for the LN electro-optic isolator.** Forward and backward spectra of the isolator are monitored using a tunable laser and 2x2 switch (forward: 1→3, backward: 1→4) to couple light onto the chip. Polarization controllers are used in both directions to couple light into the transverse-electric (TE) waveguide mode. Spectra are measured on an OSA (optical spectrum analyzer). For III-V laser operation, the DFB laser is butt-coupled directly to the isolator chip. The output is then split between a retroreflector and linewidth and relative intensity noise (RIN) measurement to monitor the behavior of the DFB laser with and without isolation.

**Characterization of DFB laser with the LN isolator.** The coupling loss between the DFB laser and the LN isolator chip is estimated to be ~2 dB based on numerical simulations and characterization measurements of the laser-LN interface. The output of the InP laser is defined by the quantum well waveguide structure drawn in Extended Data Fig. 2 and coated with anti-reflective coating designed for the quantum-well waveguide and air.

We verify the numerical results experimentally using a 4 mm LN waveguide with couplers on either side. The coupling loss between the coupler and lensed fiber is first characterized through the fiber-to-fiber loss, before the input lensed fiber is replaced by the DFB laser to measure the laser-coupler loss. The lowest coupling loss is measured to be 2 dB. There is a minimum achievable coupling gap of ~2 μm from the setup and dice angle of both chips, which agrees with the coupling simulation.

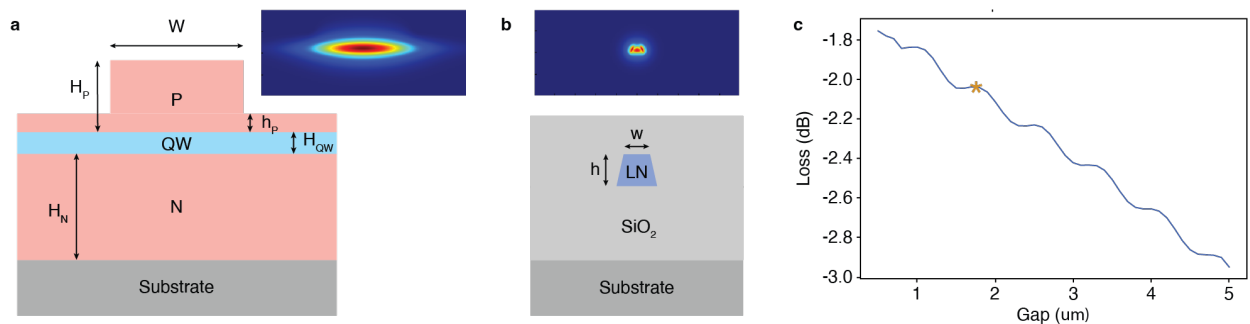

**Extended Data Fig. 2 | Coupling between DFB laser and LN chip. a,** Quantum well waveguide structure and (inset) simulated output mode of the III-V DFB laser: W = 5 um, $H_P$ = 2.5 um, $h_P$ =

500 nm, $H_{QW}$ = 450 nm, $H_N$ = 4 um; **b,** Structure of the LN mode converter at the coupling facet: h = 300 nm, w = 250 nm; **c,** Simulation of the coupling efficiency between the DFB laser mode and LN facet. An anti-reflective coating is added to the laser structure, as in experiment, to minimize reflection between the laser and the air gap. The minimum coupling gap achievable in our experiment is ~2 um, corresponding to about 2-dB coupling loss (indicated by the star). This is loss is further verified experimentally on other LN chips.

**Relative intensity noise measurement.** We sent the 10% transmission port from the DFB laser and isolator system to a photodetector followed by an RF spectral analyzer, in order to measure the relative intensity noise. Extended Data Fig. 3 shows the RIN over a 20MHz bandwidth. The RIN is not affected after the microwave driver is turned on (the isolator is on). Therefore, our isolator chip does not add intensity noise to the DFB laser when two chips are edge coupled together. However, when both isolator and reflector is turned on, we observe an increase in RIN. The cause is still under investigation since the current measurement setup does not differentiate between the noise from the laser itself and the noise induced between the reflector and chip output facet.

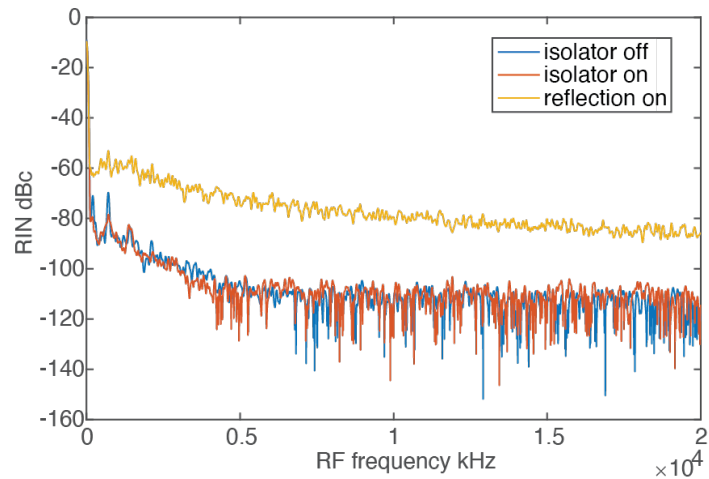

**Extended Data Fig. 3 | Relative intensity noise (RIN) measurement when isolator is turned on and off.** We observe similar RIN level between the integrated isolator turned on and off. When both reflection and isolator turned on, we observe an increase in RIN. The cause behind this phenomenon is still under investigation since the laser is followed by an extra cavity which is formed between the reflector and chip output facet.

**Table 1: Performance comparison with other on-chip isolator approaches.**

| Platform | Device | Isolation ratio (dB) | Insertion loss (dB) | Power (dBm) | Wavelength (nm) | Tested with on-chip laser | Ref |
|---|---|---|---|---|---|---|---|
| **LN (Fig.2)** | EO; waveguide | 48 | 0.5 | 21 | 1553 Tunable from 1510-1630 nm | No | **This work** |
| **LN (Fig.3)** | EO; waveguide + ring filter | 48 | 1.4 | 21 | 1552 Tunable | No | **This work** |
| **LN (Fig.4)** | EO; waveguide + ring filter | 42 | 2.9 | 21 | 1556 | Yes | **This work** |
| Ce:YIG/Si | MO; resonator; WB | 32 | 2.3 | 10 | 1555 | No | Ref[13] |
| Ce:YIG/Si | MO; resonator; WB | 32 | 11 | 15[†] | 1550 | No | Ref[14] |
| Ce:YIG/SiN | MO; resonator; PLD | 28 | 1 | NA; magnetic field | 1570 | No | Ref[12] |
| Ce:YIG/Si | MO; waveguide; PLD | 30 | 5 | NA; magnetic field | 1585 | No | Ref[15] |
| AlN/Si | AO; waveguide | 12 | 0.6 | 20.8 | 1523 | No | Ref[26] |
| LN | AO; resonator | 12.75 39.31 | 1.13 0.65 | 29 | 1550 1550 | No | Ref[25] |
| AlN/SiN | AO; resonator | 9.3 41 | 0.8 1.9 | 25 33 | 1550 1550 | No | Ref[24] |
| LN | $\chi^{(2)}$; waveguide | 18 | 6.6/14* | 5[‡] | 1575 | No | Ref[30] |
| Si | Kerr; waveguide | 4 | NA | NA | 1582 | No | Ref[29] |
| SiN | Kerr; resonator | 23 17 | 4.6 1.3 | 20[‡^] | 1550 | Yes[~] | Ref[27] |
| Silica | Kerr; resonator | 24 | 5 | 21[‡^] | 1550 | No | Ref[28] |
| InP | EO; waveguide | 11 | 2.3 | 26[†] | 1580 | No | Ref[21] |
| Si | EO; waveguide | 3 | 11.1 | 24[†] | 1556 | No | Ref[22] |
| Si | EO; resonator | 13 | 18 | -3 | 1550 | No | Ref[19] |
| GaAs/AlGaAs | EO; waveguide | 30 | 8 | 18[†] | 1550 | No | Ref[20] |
| Si | EO; waveguide | 3 | 70 | 25 | 1555 | No | Ref[17] |

[†] Extracted
*Including 3.7 dB loss per facet
[‡] Optical power
[^] Power dependent
[~] Only backward transmission is tested

### Acknowledgements
This work is supported by the Defense Advanced Research Projects Agency (HR0011-20-C-0137), ONR (N00014-18-C-1043 and N00014-22-C-1041), and AFOSR (FA9550-19-1-0376). E.P. acknowledges support by Draper graduate student fellowship. Device fabrication was performed at the Harvard University Center for Nanoscale Systems.

### Author statements
M.L. conceived the idea. M. Y. designed the chip with the help of R.C., C.R., L. H. and M. Z. C. R., K. L. and L. H. fabricated the devices. M.Y. and R. C. carried out the measurement and analyzed the data with the help from E. P., L. S. and A. S.. M. Y. performed numerical simulations. H. G. and L. J. provided the DFB laser. M. Y. and R. C. wrote the manuscript with contribution from all authors. M.L. supervised the project.

### Competing interests